\documentstyle[preprint,prl,tighten,aps,epsfig]{revtex}    %
\begin{document}

\newcommand{\wt}{\widetilde}
\newcommand{\imag}{\Im {\rm m}}
\newcommand{\real}{\Re {\rm e}}
\newcommand{\tanb}{\tan \! \beta}
\newcommand{\cotb}{\cot \! \beta}
\newcommand{\mto}{m^2_{\tilde{t}_1}}
\newcommand{\mttt}{m^2_{\tilde{t}_2}}
\newcommand{\mbo}{m^2_{\tilde{b}_1}}
\newcommand{\mbt}{m^2_{\tilde{b}_2}}
\newcommand{\ghat}{\hat{g}^2}
\newcommand{\htop}{\left| h_t \right|^2}
\newcommand{\hb}{\left| h_b \right|^2}

\draft

\preprint{\begin{tabular}{l}
\hbox to\hsize{\mbox{ }\hfill KEK--TH--754}\\%[-3mm]
\hbox to\hsize{\mbox{ }\hfill hep--ph/0103294}\\%[-3mm] 
\hbox to\hsize{\mbox{ }\hfill \today}\\%[-3mm]
\hbox to\hsize{\mbox{ }\hfill { } }\\%[-3mm]
\hbox to\hsize{\mbox{ }\hfill { } }\\%[-3mm]
          \end{tabular} }

\title{Higgs Boson Decays in the Minimal Supersymmetric Standard Model with
       Radiative Higgs Sector CP Violation}

\author{S.Y.~Choi$^1$, Kaoru Hagiwara$^2$ and 
        Jae Sik Lee$^2$ \\ \vspace{0.3 cm}} 

\address{$^1$ Department of Physics, Chonbuk National University, Chonju
         561--756, Korea \\
         $^2$ Theory Group, KEK, Tsukuba, Ibaraki 305--0801, Japan}

\maketitle

\vskip 2cm
\begin{abstract}
We re--evaluate the decays of the Higgs bosons in the minimal supersymmetric
standard model (MSSM) where the tree-level CP invariance of the Higgs 
potential is explicitly broken by the loop effects of the third--generation 
squarks with CP--violating soft--breaking Yukawa interactions. 
This study is based on the mass matrix of the neutral Higgs bosons that is 
valid for arbitrary values of all the relevant MSSM parameters. 
It extends the previous work considerably by including neutral Higgs--boson 
decays into virtual gauge bosons and those into top--squark pairs, 
by implementing squark--loop contributions to the two--gluon decay channel, 
and by incorporating the decays of the charged Higgs boson.  
The constraints from the electron electric dipole moment on the CP
phases are also discussed. We find that the branching fractions of both
the neutral and charged Higgs--boson decays and their total decay widths
depend strongly on the CP phases of the top (and bottom) squark sectors
through the loop--induced neutral Higgs boson mixing as well as the direct
couplings of the neutral Higgs bosons to top squark pairs.
\end{abstract}

\vskip 0.4cm


% 14.80.Cp Non-standard-model Higgs bosons
% 11.30.Er Charge conjugation, parity, time reversal, 
%          and other discrete symmetries
% 12.60.Jv Supersymmetric models (see also 04.65 Supergravity)

% \begin{multicols}{2}

%%%%%%%%%%%%%%%%%%%%%%%%%%%%%%%%%%%%%%%%%%%%
\section{Introduction}
\label{sec:introduction}
%%%%%%%%%%%%%%%%%%%%%%%%%%%%%%%%%%%%%%%%%%%%

The soft CP violating Yukawa
interactions in the minimal supersymmetric standard model (MSSM) cause
the CP--even and CP--odd neutral Higgs bosons to mix
via loop corrections \cite{AP,DEM,PW,CDL,CEPW}.
Although the mixing is a radiative effect, the induced CP violation
in the MSSM Higgs sector can be large enough to affect the
Higgs phenomenology significantly at present and future colliders
~\cite{AP,PW,CEPW,EXCP_FC,CL1,CL2,CL3,Bae,CL4}. \\

In the light of the possible large CP--violating mixing, we have studied in
Ref.~\cite{CL1} all the dominant two--body decay branching fractions of the 
three neutral Higgs bosons based on the mass matrix derived by Pilaftsis and
Wagner \cite{PW}. The mass matrix, however, is not applicable 
for large squark mass splitting.
In the present work, we re--evaluate all the two--body decay modes of the
neutral Higgs bosons with the newly--calculated mass matrix \cite{CDL} which
is valid for any values of the soft--breaking parameters.
Furthermore, we extend the work significantly by including the Higgs--boson
decay modes containing virtual gauge bosons and those into squark pairs and,
also by taking into account the squark--loop contributions to the 
gluon--gluon decay modes.  In addition, we study the dominant decay modes of 
the charged Higgs boson in
the presence of the non--trivial CP--violating mixing. \\

This paper is organized as follows. In Sec.~\ref{sec:CP
violation} we give a brief review of the calculation \cite{CDL}
of the loop--induced CP--violating mass matrix of
the three neutral Higgs bosons.  We also consider the 
constraints by the electron electric dipole moment (EDM)
on the space of the relevant supersymmetric parameters.
In Sec.~\ref{sec:decay} we discuss the effects of the CP
phases on the neutral Higgs boson decays. The effects of the CP
phases on the charged Higgs boson decays are discussed in Sec. \ref{sec:chdcy}.
Finally, we summarize our findings in Sec.~\ref{sec:conclude}.

%%%%%%%%%%%%%%%%%%%%%%%%%%%%%%%%%%%%%%%%%%%%%%%%%%
\section{CP Violation in the MSSM Higgs Sector}
\label{sec:CP violation}
%%%%%%%%%%%%%%%%%%%%%%%%%%%%%%%%%%%%%%%%%%%%%%%%%%

The loop--corrected mass matrix of the neutral Higgs bosons in the MSSM
can be calculated from the effective potential \cite{CW,OKADA}
\begin{eqnarray}
\label{e2}
V_{\rm Higgs}\hskip -0.3cm 
   &= \, \frac{1}{2}m_1^2\left(\phi_1^2+a_1^2\right)
     +\frac{1}{2}m_2^2\left(\phi_2^2+a_2^2\right)
     -\left|m_{12}^2\right|\left(\phi_1\phi_2-a_1 a_2\right) 
           \cos (\xi + \theta_{12}) \nonumber \\ 
   & +\left|m^2_{12}\right|
      \left(\phi_1 a_2 +\phi_2 a_1\right)\sin(\xi+\theta_{12})
     +\frac{\ghat}{8} {\cal D}^2 
     +\frac{1}{64\,\pi^2} {\rm Str} \left[
           {\cal M}^4 \left(\log\frac{{\cal M}^2}{Q^2} 
	                  - \frac{3}{2}\right)\right] \,,
\end{eqnarray}
with ${\cal D} = \phi_2^2 + a_2^2 - \phi_1^2 - a_1^2$,
$\ghat = (g^2+g'^2)/4$,
and $\phi_i$ and $a_i$ ($i=1,2$) are the real fields of the neutral
components of the two Higgs doublets:
\begin{eqnarray}
\label{e1}
H_1^0 = \frac{1} {\sqrt{2}} \left( \phi_1 + i a_1 \right)\,, \ \ \ \ \
H_2^0 = \frac {{\rm e}^{i \xi}} {\sqrt{2}} \left( \phi_2 + i a_2 \right)\,.
\end{eqnarray}
The parameters $g$ and $g'$ are the SU(2)$_L$ and U(1)$_Y$
gauge couplings, respectively, and  $Q$ denotes the
renormalization scale. All the tree--level parameters of the effective
potential (\ref{e2}) such as $m_1^2, \ m_2^2 $ and 
$m_{12}^2=\left|m^2_{12}\right|{\rm e}^{i\theta_{12}}$, are the
running parameters evaluated at the scale $Q$. The potential (\ref{e2}) is
then almost independent of $Q$ up to two--loop--order corrections. 
The super--trace is to be taken over all the bosons and fermions 
that couple to the Higgs fields.\\

The matrix ${\cal M}$ in Eq.~(\ref{e2}) is 
the field--dependent mass matrix of all modes that
couple to the Higgs bosons. The dominant contributions in the MSSM come from
third generation quarks and squarks because of their large Yukawa couplings.
The field--dependent masses of the third generation quarks are given by
\begin{eqnarray}
\label{e3}
m_b^2 = \frac{1}{2} |h_b|^2 \left( \phi_1^2 + a_1^2
\right)\,, \ \ \ \
m_t^2 = \frac{1}{2} |h_t|^2 \left( \phi_2^2 + a_2^2
\right), 
\end{eqnarray}
where $h_b$ and $h_t$ are the bottom and top Yukawa couplings, respectively. 
The corresponding squark mass matrices read:
\begin{eqnarray}
{\cal M}_{\tilde t}^2 &= \mbox{$ \left( \begin{array}{cc} 
m^2_{\wt Q} + m_t^2 - \frac{1}{8} \left( g^2 - \frac{g'^2}{3} \right) {\cal D}
\,\,\,\,\,\,\,\,\,\,
&
- h_t^* \left[ A_t^* \left(H_2^0 \right)^* + \mu H_1^0 \right] \\
- h_t \left[ A_t H^0_2 + \mu^* \left( H_1^0 \right)^* \right] &
m^2_{\wt U} + m_t^2 - \frac{g'^2}{6} {\cal D}
\end{array} \right)\,, $} \nonumber\\
\nonumber \\ 
{\cal M}_{\tilde b}^2 &= \mbox{$ \left( \begin{array}{cc} 
m^2_{\wt Q} + m_b^2 + \frac{1}{8} \left( g^2 + \frac{g'^2}{3} \right) {\cal D}
\,\,\,\,\,\,\,\,\,\,
&
- h_b^* \left[ A_b^*  \left( H_1^0 \right)^* + \mu H_2^0 \right] \\
- h_b \left[ A_b H_1^0 + \mu^* \left( H_2^0 \right)^* \right] &
m^2_{\wt D} + m_b^2 + \frac{g'^2}{12} {\cal D}
\end{array} \right)\,. $} 
\label{e4}
\end{eqnarray}
where $m^2_{\wt Q}, \ m^2_{\wt U}$ and
$m^2_{\wt D}$ are the real soft SUSY--breaking squark-mass
parameters, $A_b$ and $A_t$ are the
complex soft SUSY--breaking trilinear parameters, and $\mu$ is the complex
supersymmetric Higgsino mass parameter.\\

The mass matrix of the Higgs bosons (at vanishing external momenta)
is then given by the second derivatives of the potential, 
evaluated at its minimum point
\begin{eqnarray}
  \left(\phi_1,\, \phi_2,\, a_1,\, a_2\right)
=\left(\langle \phi_1\rangle,\, \langle \phi_2 \rangle,\,
       \langle a_1\rangle,\, \langle a_2 \rangle\right)
=\left(v\cos\beta,\, v\sin\beta,\, 0,\, 0\right),
\end{eqnarray}
where $v=(\sqrt{2}\, G_F)^{-1/2}\simeq 246 \ {\rm GeV}$. 
% $\langle a_i \rangle = 0$, $\langle \phi_1 \rangle^2 
% + \langle \phi_2 \rangle^2 = v^2
% \simeq (246 \ {\rm GeV})^2$ and $\langle \phi_2 \rangle/
% {\langle \phi_1 \rangle} = \tanb$.
The massless state $G^0 = a_1 \cos \beta - a_2 \sin \beta$ is
the would--be--Goldstone mode to be absorbed by the $Z$ boson. 
We are thus left with a mass--squared matrix ${\cal M}_H^2$ for three 
physical states, 
$a \,(= a_1 \sin \beta + a_2 \cos \beta), \ \phi_1$ and $\phi_2$. 
This matrix is real and symmetric, i.e. it has 6 independent entries. 
The diagonal entry for the pseudoscalar component $a$ reads:
\begin{eqnarray}
\label{e10}
\left. {\cal M}^2_{H} \right|_{aa} = m_A^2 + \frac {3} {8 \pi^2}
\left\{ \frac { |h_t|^2 m_t^2 } { \sin^2 \beta} g(\mto,
\mttt) \Delta_{\tilde t}^2 +
\frac {|h_b|^2 m_b^2 } { \cos^2 \beta} g(\mbo,
\mbt) \Delta_{\tilde b}^2 \right\}\,,
\end{eqnarray}
where $m_A$ is the loop--corrected pseudoscalar mass in the CP invariant
theories.
The CP--violating entries of the mass matrix,
which mix $a$ with $\phi_1$ and $\phi_2$, are given by
\begin{eqnarray}
\label{e13}
\left. {\cal M}^2_H \right|_{a \phi_1}
   &=& \frac {3} {16 \pi^2} \left\{
       \frac { m_t^2 \Delta_{\tilde t} } {\sin \beta} \left[ g(\mto, \mttt)
       \left( X_t \cotb - 2 \htop R_t \right)
            - \ghat \cotb \log \frac{\mttt}{\mto} \right] \right. \\
   && \left. \hskip 1cm
     +\frac {m_b^2 \Delta_{\tilde b}} {\cos \beta} \left[ -g(\mbo,\mbt)
      \left( X_b + 2 \hb R_b' \right) + \left( \ghat - 2 \hb \right) \log
      \frac {\mbt} {\mbo} \right] \right\}, \nonumber\\
\left. {\cal M}^2_H \right|_{a \phi_2}
   &=& \frac {3} {16 \pi^2} \left\{
       \frac {m_t^2 \Delta_{\tilde t}} {\sin \beta} \left[ -g(\mto,\mttt)
       \left( X_t + 2 \htop R_t' \right)
            + \left(\ghat-2\htop\right)\log\frac{\mttt}{\mto} \right]\right.
\\
   && \left. \hskip 1cm
     +\frac { m_b^2 \Delta_{\tilde b} } {\cos \beta} \left[ g(\mbo, \mbt)
      \left( X_b \tanb - 2 \hb R_b \right)
            - \ghat\tanb\log \frac{\mbt}{\mbo} \right]\right\}. \nonumber
\end{eqnarray}
where $g(x,y)=2-[(x+y)/(x-y)]\log(x/y)$.
The size of these CP--violating entries is determined by
the re--phasing invariant quantities
%$\Delta_{\tilde t}$ and $\Delta_{\tilde b}$ given by
\begin{eqnarray}
\label{e9}
\Delta_{\tilde t} = \frac { \imag(A_t \mu {\rm e}^{i \xi}) }
{\mttt - \mto} \,, \ \ \qquad
\Delta_{\tilde b} = \frac { \imag(A_b \mu {\rm e}^{i \xi}) }
{\mbt - \mbo} ,
\end{eqnarray}
which measure the amount of CP violation in the top and bottom squark--mass
matrices. In the CP--conserving limit, both  
$\Delta_{\tilde t}$ and $ \Delta_{\tilde b}$ vanish, leading to
$|m^2_{12}|\,\sin(\xi + \theta_{12})=0$.
The definition of the mass--squared $m^2_A$ and the dimensionless quantities
$X_{t,b}$, $R_{t,b}$ and $R^\prime_{t,b}$,
as well as the other CP--preserving entries of
the mass matrix squared ${\cal M}^2_{H}$,  can be found in Ref.~\cite{CDL}. 
The real and symmetric matrix ${\cal M}^2_H$ can now be
diagonalized with an orthogonal matrix $O$;
\begin{eqnarray}
\label{OMIX}
%O^T\,{\cal M}^2_H\, O={\rm diag}(m^2_{H_1},m^2_{H_2},m^2_{H_3}),
\left(\begin{array}{c}
       a      \\
       \phi_1 \\
       \phi_2
      \end{array}\right)\,=\,O\,\,
\left(\begin{array}{c}
       H_1 \\
       H_2 \\
       H_3
      \end{array}\right)\,.
\end{eqnarray}
Our convention for the three mass eigenvalues is
$m_{H_1}\leq m_{H_2}\leq m_{H_3}$. \\

The loop--corrected neutral--Higgs--boson sector depends on
various parameters from the other sectors of the MSSM; $m_A$, $\tan\beta$, 
$\mu$, $A_t$, $A_b$, the renormalization scale $Q$, 
and the real soft--breaking masses, $m_{\tilde Q}$, 
$m_{\tilde U}$, and $m_{\tilde D}$, as well as on the complex gluino--mass
parameter $M_{\wt g}$ through one--loop corrections to the top and bottom
quark masses \cite{SUSYHBB}.  
Noting that the size of the radiative Higgs sector CP violation is determined
by the rephasing invariant combinations $A_t \mu {\rm e}^{i\xi}$ and 
$A_b \mu {\rm e}^{i\xi}$, see Eq. (\ref{e9}), 
we take for our numerical analysis the following set of parameters:
\begin{eqnarray}
&& |A_t|=|A_b| = 1~{\rm TeV}\,, 
\hspace{1.9 cm} |\mu|=2~{\rm TeV}\,,\nonumber \\
&& M_{{\widetilde Q},{\widetilde U},{\widetilde D}}=|M_{\widetilde g}|
= 0.5~{\rm TeV}\,,
\qquad \xi+{\rm Arg}(\mu)={\rm Arg}(M_{\widetilde g})=0\,,
\label{eq:PARA}
\end{eqnarray}
under the constraint;
\begin{equation}
\Phi \equiv {\rm Arg}(A_t\mu {\rm e}^{i\xi}) = 
{\rm Arg}(A_b\mu {\rm e}^{i\xi})\,.
\end{equation}
We vary the common phase $\Phi$ as well as $m_A$ and $\tan\beta$ in 
the following numerical studies. Our choice of relatively large magnitudes of
$|A_t\mu|=|A_b\mu|$ enhances CP--violation effects in the MSSM Higgs sector
\footnote{
Our choice of the set of parameters satisfies the necessary condition to avoid 
a color and electric--charge breaking (CCB) minimum
in the direction $|{\widetilde Q}|=|{\widetilde U}|=|H^0_2|$ \cite{CCB-OLD}:
$ |A_t|^2 \leq 3(M^2_{{\widetilde Q}}+M^2_{{\widetilde U}}+m_2^2) $.
But, according to the more general study \cite{CLM},
our relatively large values of $|A_{t,b}|$ and $|\mu|$ compared to
those of $M_{{\widetilde Q},{\widetilde U},{\widetilde D}}$ could give rise
to a dangerous CCB minima in the potential which could be 
deeper than the electro--weak vacuum. Therefore, a detailed study of the 
CCB minima in the presence of the non--trivial CP--violating mixing among the
neutral Higgs bosons deserves further analysis. 
}.
\\

Before studying the CP--violation effects on the neutral and 
charged Higgs--boson decays,
it is worthwhile to examine the present experimental
constraints on the above parameter values (\ref{eq:PARA}). 
The charginos, neutralinos, and squarks of the first two generations 
are sufficiently heavy for the parameter set (\ref{eq:PARA}). 
The lightest squark is the lighter top squark
$\tilde{t}_1$, the mass of which can be as low as 200 GeV for $\tan\beta=4$ and
$\Phi=0^{\rm o}$. The CP--violating phase could weaken the LEP lower 
limit on the lightest Higgs boson mass significantly \cite{KW,CEPW2}. 
In our analysis we show our results when the lightest Higgs--boson mass is
above 70 GeV. \\

\begin{figure}
\begin{center}
\hbox to\textwidth{\hss\epsfig{file=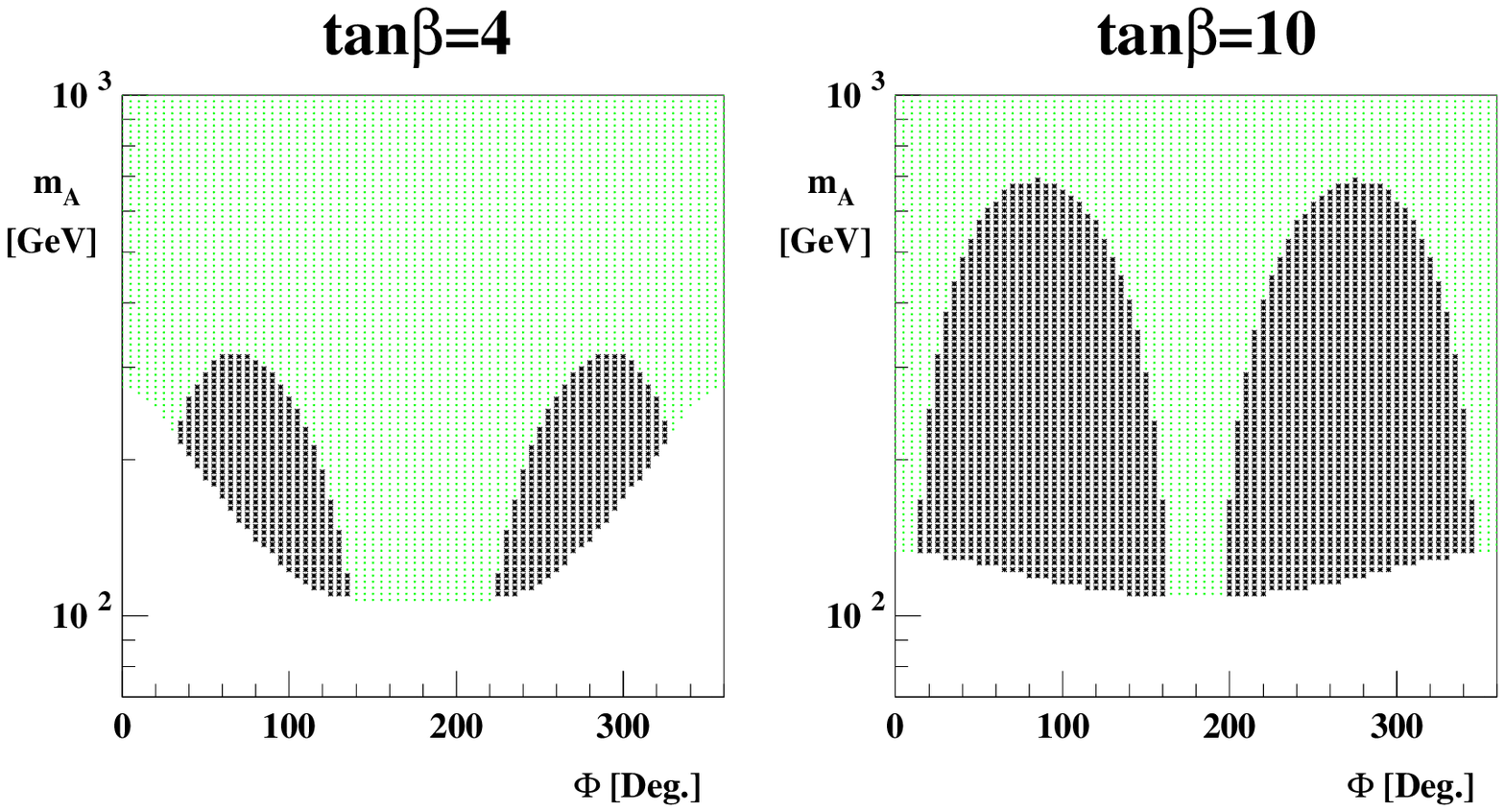,width=16cm,height=16cm}\hss}
\end{center}
\caption{The regions of $m_A$ and $\Phi$ allowed by the experimental
         electron EDM constraint for the parameter set (\ref{eq:PARA}) 
	 are lightly shaded for both $\tan\beta=4$ (left) and 
	 $\tan\beta=10$ (right). The dark--shaded regions are excluded 
	 by the electron EDM constraint while the unshaded regions give
	 $m_{H_1} < 70$ GeV.}
\label{edm}
\end{figure}

The one--loop effective couplings of the CP--odd components of the Higgs 
boson to the gauge bosons give rise to the electron and neutron EDM's
\cite{CKP} at two--loop level. The effects could be significant for large
$\tan\beta$ at large $|A_t|,|A_b|$ and $|\mu|$ so that some region of our
parameter space is already excluded by the present $2\sigma$ upper bounds 
on the electron and neutron EDM's : $|d_e|<0.5\times 10^{-26}\, e\,$cm and 
$|d_n|<1.12\times 10^{-25}\, e\,$cm, respectively \cite{PDG}.  
The dark--shaded region in Fig.~\ref{edm} is excluded by the electron 
EDM constraint in the ($\Phi$,\, $m_A$) plane for $\tan\beta=4$ and $10$. 
The unshaded regions give $m_{H_1} < 70$ GeV. The EDM constraints are 
avoided only :n the lightly shaded region. The excluded region becomes 
larger for larger $\tan\beta$. Even for $\tan\beta=4$ as shown in 
Fig.~\ref{edm}, some parameter space with small $m_A$  and large
CP--violating phase can be excluded by the EDM constraint. \\

It should be noted, however, that the strong two--loop EDM constraints 
due to the one--loop effective couplings of the CP--odd components of the
Higgs bosons to the gauge bosons may not be valid if there appears 
cancellation among different EDM contributions.
Such cancellations may occur between one-- and two--loop contributions, or 
among two--loop contributions themselves \cite{CEPW2,P-EDM}. 
In fact, the excluded regions of
Fig.~\ref{edm} disappear if such cancellation takes place at  
50~\% (20~\%) level for $\tan\beta=4$ ($\tan\beta=10$).
In this paper, we show our results for the whole parameter space
(\ref{eq:PARA}).

%
% Pilaftsis's e-mail
%
% This is true. Therefore, for the CP-violating MSSM benchmark scenario we
% were recently advocating in hep-ph/0009212 and in some of our earlier
% papers, you may need order-one (mild) cancellations even for low values of
% tan(beta). On the other hand, you need cancellations at the 10% level for
% tan(beta) > 20. Such cancellations may occur between one- and two-loop
% graphs or between two-loop EDM's only (please see also remarks in
% hep-ph/9912253).  In any case, the degree of cancellations is not as
% dramatic as the one required for the one-loop EDM's for light 1st
% generations squarks, where it is less than 1% for M(1gen. squarks) < 200
% GeV.
%

%%%%%%%%%%%%%%%%%%%%%%%%%%%%%%%%%
\section{Neutral Higgs Boson Decays}
\label{sec:decay}
%%%%%%%%%%%%%%%%%%%%%%%%%%%%%%%%%

In this section we discuss the phenomenological consequences
of the CP--violating Higgs--boson mixing on
the total decay widths and decay branching fractions
of the neutral Higgs bosons. We choose
$\tan\beta=4$ throughout this section,
which leads larger CP--violating effects than the $\tan\beta=10$ case
\cite{CL1}. \\

The most important decay channels of the Higgs bosons are 
two--body decays into the heaviest fermions and bosons
because the Higgs couplings
are proportional to the particle masses. We refer to
Ref.~\cite{CL1} for explicit forms of the two--body decay widths of each Higgs
boson and the relevant interaction Lagrangians. \\

The Higgs boson decays into virtual gauge bosons $V^{(*)}V^{(*)}$ 
%followed by their leptonic decays 
are also important for the Higgs bosons of the intermediate mass region,
$110$ GeV $\lesssim m_H \lesssim$ $150$ GeV \cite{KM,DKZ,THAN}. 
Because of the importance of these 3--body and 4--body decay modes, we extend
the previous work \cite{CL1}
to include the Higgs--boson decays into virtual gauge
boson pairs and those into a lighter Higgs boson and a virtual $Z$ boson.
The partial decay width of $H_i\rightarrow V^{(*)} V^{(*)}$
is given by
\begin{eqnarray}
&&\Gamma(H_i\rightarrow V^{(*)} V^{(*)})=
\frac{G_F m^3_{H_i}\delta_V(\cos\beta O_{2i}+\sin\beta O_{3i})^2}
{16\sqrt{2}\pi}\nonumber \\
&\times&\int_0^{\omega_i} {\rm d}x\int_0^{(\sqrt{\omega_i}-\sqrt{x})^2}{\rm d}y
\frac{\epsilon^2_{_V}\lambda^{1/2}(\omega_i,x,y)
(\lambda(\omega_i,x,y)+12xy)}
{\omega_i^3\pi^2[(x-1)^2+\epsilon^2_{_V}][(y-1)^2+\epsilon^2_{_V}]} \,,
\end{eqnarray}
where $\epsilon_{_V}=\Gamma_V/m_V$,
$\omega_i=m_{H_i}^2/m_V^2$, $\delta_W=2$, $\delta_Z=1$, and
$\lambda(x,y,z)=x^2+y^2+z^2-2xy-2yz-2zx$.
For $m_{H_i}>2 m_V$, $1/[(x-1)^2+\epsilon^2]$ can be approximated by
$\pi/\epsilon\,\delta(x-1)$ 
and the expression of Ref.~\cite{CL1} is recovered for small $\epsilon$.
On the other hand, the partial decay width of $H_i\rightarrow H_j Z^{(*)}$
is given by
%\begin{eqnarray}
%\Gamma(H_i\rightarrow H_j Z^{(*)})&=&
%\frac{G_F m^3_{H_i}\left[O_{1i}(\cos\beta O_{3j}-\sin\beta O_{2j})
%-(i\leftrightarrow j)\right]^2} {8\sqrt{2}\pi} \nonumber \\
%&&\times
%\int_0^{(\sqrt{\omega_i}-\sqrt{\omega_j})^2}{\rm d}x\,
%\frac{{\epsilon_Z \lambda^{3/2}(\omega_i,\omega_j,x)}}
%{\omega^3_i\pi[(x-1)^2+\epsilon^2_Z]} \,.
%\end{eqnarray}
\begin{equation}
\Gamma(H_i\rightarrow H_j Z^{(*)})=
\frac{G_F m^3_{H_i} g_{_{H_iH_jZ}}^2} {8\sqrt{2}\pi} 
\int_0^{(\sqrt{\omega_i}-\sqrt{\omega_j})^2}{\rm d}x\,
\frac{{\epsilon_Z \lambda^{3/2}(\omega_i,\omega_j,x)}}
{\omega^3_i\pi[(x-1)^2+\epsilon^2_Z]} \,,
\end{equation}
where
$ g_{_{H_iH_jZ}}=O_{1i}(\cos\beta\,O_{3j}-\sin\beta\,O_{2j})
-O_{1j}(\cos\beta\,O_{3i}-\sin\beta\,O_{2i})
$.  \\

For the parameter set (\ref{eq:PARA}), the decays of two heavy Higgs bosons 
into sfermion pairs are kinematically allowed and their decay widths are 
given by
\begin{equation}
\Gamma(H_i\rightarrow \tilde{f}_j^*\tilde{f}_k)=\frac{N_C}{16\pi m_{H_i}}
|g^i_{\tilde{f}_j\tilde{f}_k}|^2
\lambda^{1/2}(1,m_{\tilde{f}_j}^2/m_{H_i}^2,m_{\tilde{f}_k}^2/m_{H_i}^2)\,,
\end{equation}
where $N_C=3(1)$ for squarks (sleptons).
Finally, the Higgs boson decay width into 
a gluon pair including the sfermion contribution to the decay width
is given by
\begin{equation}
\Gamma(H_i\rightarrow g\,g)=\alpha_s^2\frac{m_{H_i}}{32\pi^3}
\left(|S^g_i(m_{H_i}^2)|^2+|P^g_i(m_{H_i}^2)|^2\right) \,,
\end{equation}
where the dimensionless form factors $S^g_i(s)$ and $P^g_i(s)$ 
are given by
\begin{eqnarray}
S^g_i(s)&=&\sum_{f=t,b}\left\{
g_{sff}^i\frac{\sqrt{s}}{m_f} F_{sf}\left(\frac{s}{4m_f^2}\right)
+\frac{1}{4}\sum_{j=1,2}
g_{\tilde{f}_j\tilde{f}_j}^i
\frac{\sqrt{s}}{m_{\tilde{f}_j}^2} F_0\left(\frac{s}{4m_{\tilde{f}}^2}\right)\right\}
\,, \nonumber \\
P^g_i(s)&=&\sum_{f=t,b} g_{pff}^i\frac{\sqrt{s}}{m_f}
F_{pf}\left(\frac{s}{4m_f^2}\right) \,.
\end{eqnarray}
The couplings $g_{sff}^i$, $g_{pff}^i$, and $g_{\tilde{f}_j\tilde{f}_j}^i$ and
the functions $F_{sf}$, $F_{pf}$, and $F_0$ are given, for example, 
in \cite{CL3}. \\

%describe the
%loop--induced interaction of the neutral Higgs bosons with a gluon pair as:
%\begin{eqnarray} \label{hgg}
%&&\sqrt{s}\frac{\alpha_s \delta^{ab}}{4\pi}\bigg\{S^g_i(s)
%\left(\epsilon_1\cdot\epsilon_2-\frac{2}{s}k_1\cdot\epsilon_2
%\,k_2\cdot\epsilon_1\right)
% -P^g_i(s)\frac{2}{s}
%\epsilon_{\mu\nu\rho\sigma}\,\epsilon_1^\mu \epsilon_2^\nu k_1^\rho
%k_2^\sigma\bigg\}\,,
%\end{eqnarray}
%where $a$ and $b$ ($a,b=1$ to $8$) are indices of the eight SU(3) generators
%in the adjoint representation,
%$k_{1,2}$ are the four--momenta of the two gluons and $s=(k_1+k_2)^2$.
%For the decay of Higgs boson $H_i$, $\sqrt{s}$ is to be replaced with $m_{H_i}$.
%The wave vectors of the two gluons are denoted by $\epsilon_{1,2}$. The
%$s$--dependent dimensionless form factors
%$S^g_i(s)$ and $P^g_i(s)$ 

%
%The QCD radiative corrections \cite{SDGZ} to the two--gluon channel may
%be sizable and are built up by the exchange of virtual gluons, gluon radiation
%from the internal quark loop and the splitting of a gluon into two
%unresolved gluons or quark--antiquark pair. Although they are large
%enough to nearly double the partial width, we neglect the corrections
%in our numerical analyses because after all the two--gluon channel
%remains as a sub--dominant decay channel for all the neutral Higgs
%bosons.

\begin{figure}
\begin{center}
\hbox to\textwidth{\hss\epsfig{file=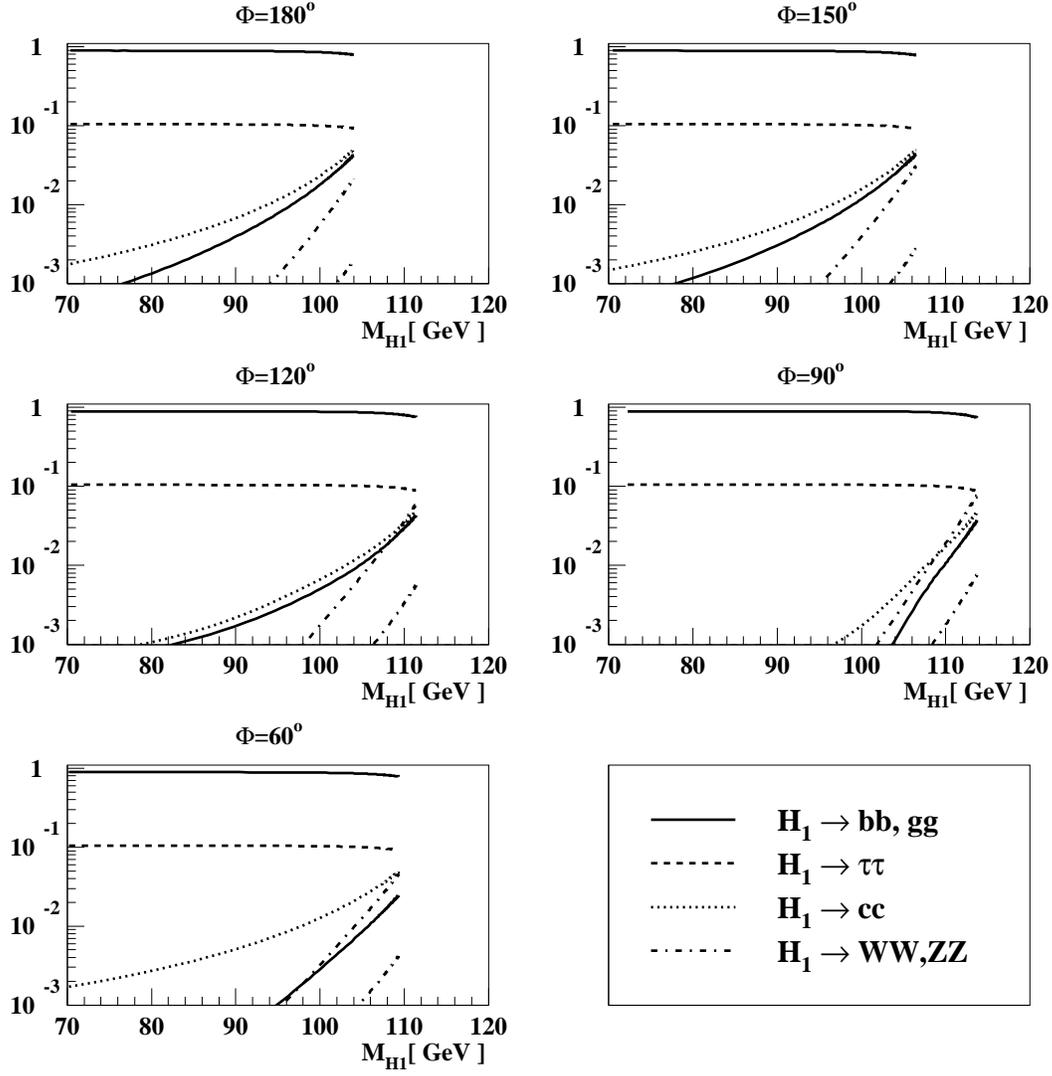,width=16cm,height=16cm}\hss}
\end{center}
\caption{The partial branching fractions of the lightest Higgs--boson decays 
         with respect to the mass $m_{H_1}$ for $\tan\beta=4$ and five values 
	 values of the CP phase $\Phi$ ; 
	 $\Phi=180^{\rm o}\,,150^{\rm o}\,,120^{\rm o}\,,90^{\rm o}\,,$ and 
         $60^{\rm o}$. The upper solid line in each frame is for 
	 $H_1\rightarrow b\bar{b}$, the dashed line for $H_1\rightarrow 
	 \tau^+\tau^-$, the dotted line for $H_1\rightarrow c\bar{c}$, 
         and lower solid line for $H_1\rightarrow gg$. 
         In addition, the upper dash--dotted line is for $H_1\rightarrow 
	 W^{(*)}W^{(*)}$ and the lower one for 
	 $H_1\rightarrow Z^{(*)}Z^{(*)}$.}
\label{br1}
\end{figure}

Figure~\ref{br1} shows the dominant partial branching fractions of the 
lightest Higgs boson decays with respect to the mass $m_{H_1}$ for five
different values of the CP phase $\Phi$ :
$\Phi=180^{\rm o}$, $150^{\rm o}$, $120^{\rm o}$,
$90^{\rm o}$, and $60^{\rm o}$. 
The lightest Higgs boson mass is found to be
less than 120 GeV and, as a result,
the main decay channels are $b\bar{b}$ (upper solid line), 
$\tau^+\tau^-$ (dashed line), $c\bar{c}$ (dotted line), and 
$gg$ (lower solid line). We note
that the pattern of the branching fractions of the main decay 
channels, $H_1\rightarrow b\bar{b}$ and $\tau^+\tau^-$,
is almost independent of the CP phase. However, for larger $m_{H_1}$,
the branching fraction of the decay 
$H_1\rightarrow W^{(*)} W^{(*)}$ becomes comparable to that of 
the two--gluon channel for $\Phi=120^{\rm o}$
and it can be as large as that of the $\tau^+\tau^-$
channel for $\Phi=90^{\rm o}$
\footnote{It is worthwhile to note that the decay mode $H_1 \rightarrow
\gamma \gamma$ with $m_{H_1}\leq 130$ GeV is crucial for the detection of
the lightest Higgs boson at the CERN Large Hadron Collider (LHC)
despite its very low branching fraction. A
detailed analysis of the dependence of this important decay mode on the CP
phases is to be reported elsewhere \cite{CHL}.}.
\\

\begin{figure}
\begin{center}
\hbox to\textwidth{\hss\epsfig{file=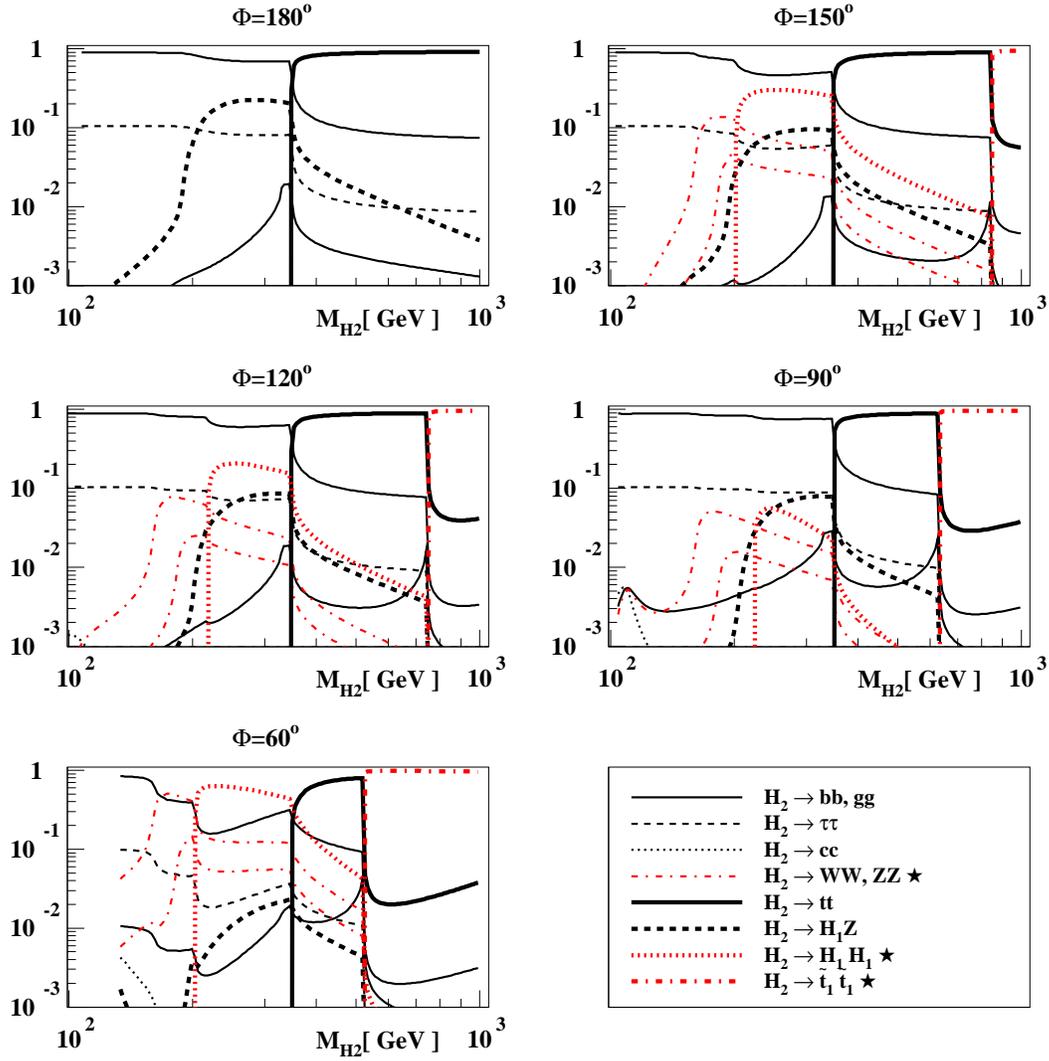,width=16cm,height=16cm}\hss}
\end{center}
\caption{The partial branching fractions for the $H_2$ decay
         channels with respect to the mass $m_{H_2}$ for five values of the
         CP phase $\Phi$ for $\tan\beta=4$ as in Fig.~\ref{br1}. 
         The decay channels marked by filled stars in the legend are forbidden 
         in the CP invariant theories.}
\label{br2}
\end{figure}

\begin{figure}
\begin{center}
\hbox to\textwidth{\hss\epsfig{file=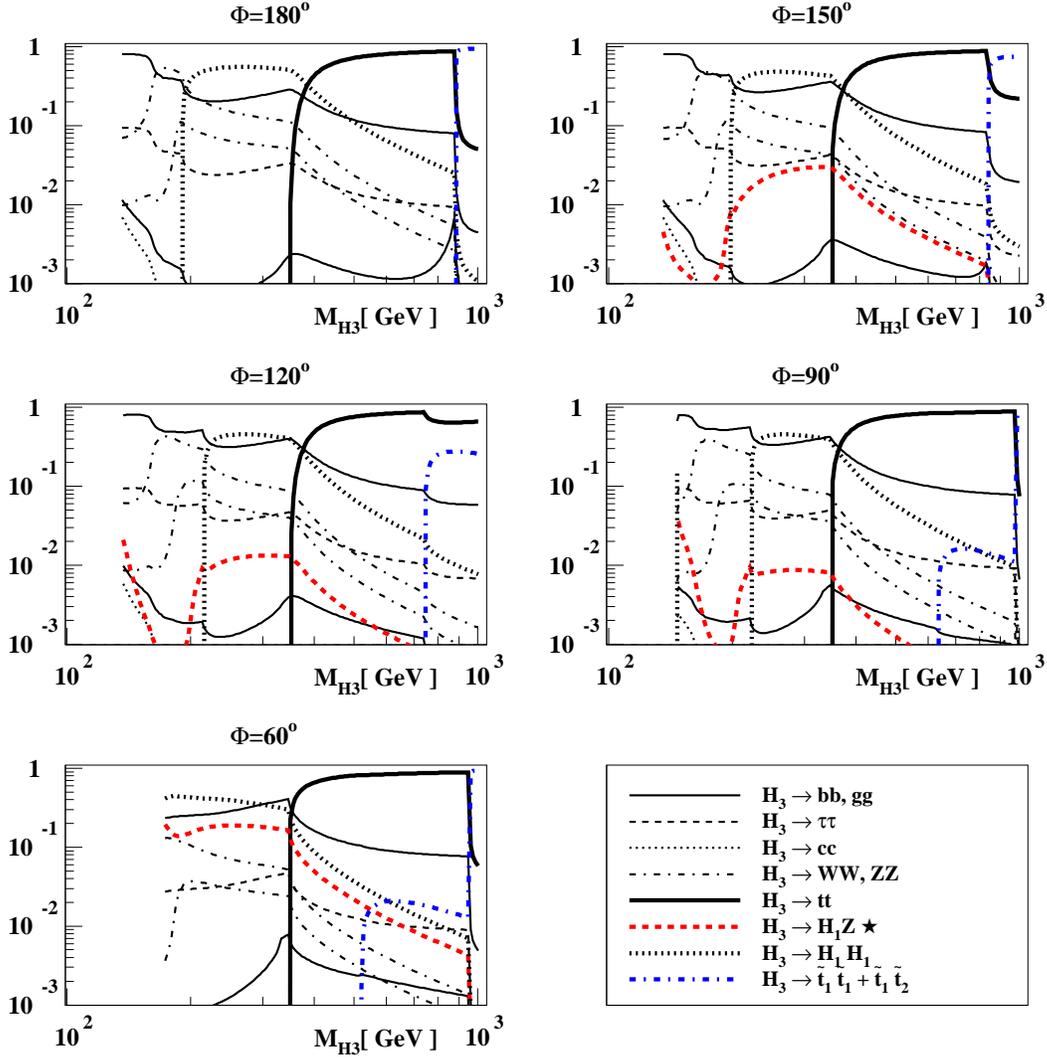,width=16cm,height=16cm}\hss}
\end{center}
\caption{The partial branching fractions for the $H_3$ decay
         channels with respect to the mass $m_{H_3}$ for $\tan\beta=4$ and
         five values of the CP phase $\Phi$ as in Figs.~\ref{br1} and 
	 \ref{br2}. The thick dash--dotted line is for the sum of the
         $H_3\rightarrow {\tilde t}_1\overline{{\tilde t}_1}$ and 
	 $H_3\rightarrow {\tilde t}_1\overline{{\tilde t}_2}
          +{\tilde t}_2\overline{{\tilde t}_1}$ modes.
         The decay channel $H_3\rightarrow H_1 Z$ marked by a filled star 
	 in the legend is forbidden in the CP invariant theories.}
\label{br3}
\end{figure}

On the other hand,
the partial branching fractions for the $H_{2,3}$ decays
are very sensitive to the CP phase $\Phi$ and strongly dependent on their
masses $m_{H_{2,3}}$ as clearly shown in Figs.~\ref{br2} and \ref{br3}. 
The upper solid line in each frame is for $H_{2,3}\rightarrow b\bar{b}$ and
the lower solid line for $H_{2,3}\rightarrow g\,g$.
The dashed line is for $H_{2,3}\rightarrow \tau^+\tau^-$,
the dotted-line for $H_{2,3}\rightarrow c\bar{c}$, and
two dash--dotted lines are for the channels 
$H_2\rightarrow W^{(*)}W^{(*)}$ (upper line)
and $Z^{(*)}Z^{(*)}$ (lower line), respectively. 
In addition, the thick solid line shows the branching fractions
for $H_{2,3}\rightarrow t\bar{t}$, 
the thick dashed line is for $H_{2,3}\rightarrow H_1 Z^{(*)}$, and
the thick dotted-line is for $H_{2,3}\rightarrow H_1\,H_1$. Finally,
the thick dash--dotted lines are for the ${\tilde t}_i\overline{{\tilde t}_j}$
decay channels.
The decay channels  forbidden in the CP invariant theories are marked by filled
stars in the legend.  These channels become significant as the phase
$\Phi$ differs from $180^{\rm o}$, being the most dominant channels of the
heavy Higgs bosons for non--trivial values of $\Phi$.
\\

Relegating the detailed descriptions of the 2--body decay channels of the
heavy Higgs bosons to Ref.~\cite{CL1}, it is worthwhile to summarize
a few interesting features of the decay modes into virtual vector
boson pairs and those into top--squark pairs :
\begin{itemize}
\item 
The decays $H_2\rightarrow W^{(*)}W^{(*)}$ 
could become the second dominant channel for $M_{H_2}\leq 2\,M_W$
for a large range of nontrivial $\Phi$.
And the branching fraction of the same decay channel for the heaviest 
Higgs boson, $H_3\rightarrow W^{(*)}W^{(*)}$, is also sizable
for the similar range of $\Phi$.
\item 
The decay channel $H_2\rightarrow{\tilde t_1}\overline{{\tilde t_1}}$,
forbidden in the CP invariant theories, overwhelms all the other decay 
channels  for the most range of $\Phi$, 
once the mode is kinematically allowed. On the contrary,
the branching fraction of the decay channel 
$H_3\rightarrow{\tilde t_1}\overline{{\tilde t_1}}$ 
decreases as $\Phi$ differs from
$180^{\rm o}$, exhibiting a typical two--state mixing between two Heavy Higgs
bosons for large $m_A$. 
\item
For the parameter set (\ref{eq:PARA}), the non--diagonal decay channels into 
$H_3 \rightarrow {\tilde t_1}\overline{{\tilde t_2}}$ and 
${\tilde t_2}\overline{{\tilde t_1}}$ dominate almost all the
other decay channels of $H_3$ once the channel is kinematically allowed 
when $\Phi$ differs significantly from $180^{\rm o}$. 
The dominance of these channels can be clearly seen near the right 
end of each frame in Fig.~\ref{br3} with $\Phi=90^{\rm o}$ and $60^{\rm o}$.
\end{itemize}
The dominance of the top--squark channels in the heavy Higgs-boson decays
is mainly due to the large values of $|A_t|$ and $|\mu|$ in Eq. (\ref{eq:PARA}),
which enhance the Higgs--boson couplings to the top squarks.
For the particular parameter set (\ref{eq:PARA}), two top squarks mix
with each other (almost) maximally, leading to a significant suppression 
of the coupling of CP--even Higgs bosons to non--diagonal top squarks.

%To recapitulate, the branching fractions of the
%lightest Higgs boson $H_1$ are not so sensitive to the CP--violating phase,
%but those of the heavy Higgs states are very sensitive to the phase.

%%%%%%%%%%%%%%%%%%%%%%%
\section{Charged Higgs Boson Decays}
\label{sec:chdcy}
%%%%%%%%%%%%%%%%%%%%%%%
Below the $H^+\rightarrow t \bar{b}$ decay threshold,
the decay channel $H^+\rightarrow W^{+(*)}H_1$, involving the lightest Higgs
boson, constitutes one of the major decay channels of the charged Higgs boson
together with the decay channel $H^+\rightarrow \tau^+\nu_\tau$. The fermionic
decay channels $H^+\rightarrow t \bar{b}$ and $\tau^+\nu_\tau$ are not
affected by the CP--violating neutral Higgs mixing. On the contrary, the
coupling of the $H^\pm W^\mp H_1$ vertices depend strongly on the mixing among
neutral Higgs bosons such that the phase $\Phi$ can affect the branching
fractions of the charged--Higgs--boson decays. \\

The interaction of the charged Higgs boson 
with a neutral Higgs boson and a $W$ boson in the presence of the CP--violating
neutral--Higgs--boson mixing is described by the Lagrangian
\begin{equation}
{\cal L}_{H^{\pm}W^{\mp}H_i}=\frac{ig}{2}\left(\sin\beta\, O_{2i}-
\cos\beta\, O_{3i} +i\,O_{1i}\right)\,
\left( H_i \stackrel{\leftrightarrow}{\partial^\mu} H^+\right)W_\mu^-
+{\rm H.c.}\,,
\end{equation}
in contrast to which the couplings of the neutral Higgs fields to 
vector bosons are determined by the Lagrangian
\begin{equation}
{\cal L}_{H_iVV}=gm_W(\cos\beta O_{2i}+\sin\beta O_{3i})
    \, H_i\left[W^+_\mu W^{-\mu}+\frac{1}{2c^2_W} Z_\mu Z^\mu\right]\,.
\end{equation}
Note that if the $H_iVV$ couplings are suppressed, 
the $H^{\pm}W^{\mp}H_i$ couplings are
enhanced and vice versa because of the orthogonality of the mixing matrix $O$:
$O_{1i}^2+O_{2i}^2+O_{3i}^2=1$. It was recently shown \cite{PW,KW} that 
the experimental lower limits from the processes 
$e^+e^-\rightarrow ZH_1$ and $e^+e^-\rightarrow \nu\bar{\nu} H_1$ on
$m_{H_1}$ and $\tan\beta$ could be significantly relaxed with the suppression
of the $H_1VV$ couplings due to the CP--violating neutral Higgs--boson mixing. 
In this case, the decay channel $H^\pm
\rightarrow W^{\pm(*)} H_1$ plays a crucial role in confirming the existence
of the CP--violating neutral--Higgs--boson mixing because this mode must be
enhanced due to the orthogonality relation. \\

The explicit forms of the two-- and three--body decay widths of the
fermionic modes, $H^+\rightarrow t^{(*)}\bar{b}, \tau^+\nu_\tau$,
can be found in Ref.~\cite{DKZ}. So, we present in the present
work just the expression of the decay width of 
$H^+\rightarrow W^{+(*)}H_i$:
\begin{equation}
\Gamma(H^+\rightarrow W^{+(*)}H_i)=
\frac{g^2 m_{H^\pm} \left|g_{_{H^\pm W^\mp H_i}}\right|^2}
{64\pi^2}
\int_0^{(1-\kappa_{_{H_i}})^2} {\rm d}x\,
\frac{\gamma_{_W}/\kappa_{_W}\,\lambda^{3/2}(1,x,\kappa^2_{_{H_i}})}
{(x-\kappa^2_{_W})^2+\kappa^2_{_W}\gamma^2_{_W}}\,,
\end{equation}
where $g_{_{H^\pm W^\mp H_i}}=\sin\beta\, O_{2i}-
\cos\beta\, O_{3i} +i\,O_{1i}$, $\kappa_x= m_x/m_{H^\pm}$, and
$\gamma_{_W}= \Gamma_{_W}/m_{H^\pm}$. For the charged Higgs--boson mass
$m_{H^\pm}\geq m_{H_i}+m_W$, the
decay width simplifies in the narrow--width approximation to
\begin{equation}
\Gamma(H^+\rightarrow W^{+}H_i)=
\frac{G_F m_{H^\pm}^3\left|g_{_{H^\pm W^\mp H_i}}\right|^2}
{8\sqrt{2}\pi}\lambda^{3/2}(1,\kappa^2_{_W},\kappa^2_{_{H_i}})\,.
\end{equation}

\begin{figure}
\begin{center}
\hbox to\textwidth{\hss\epsfig{file=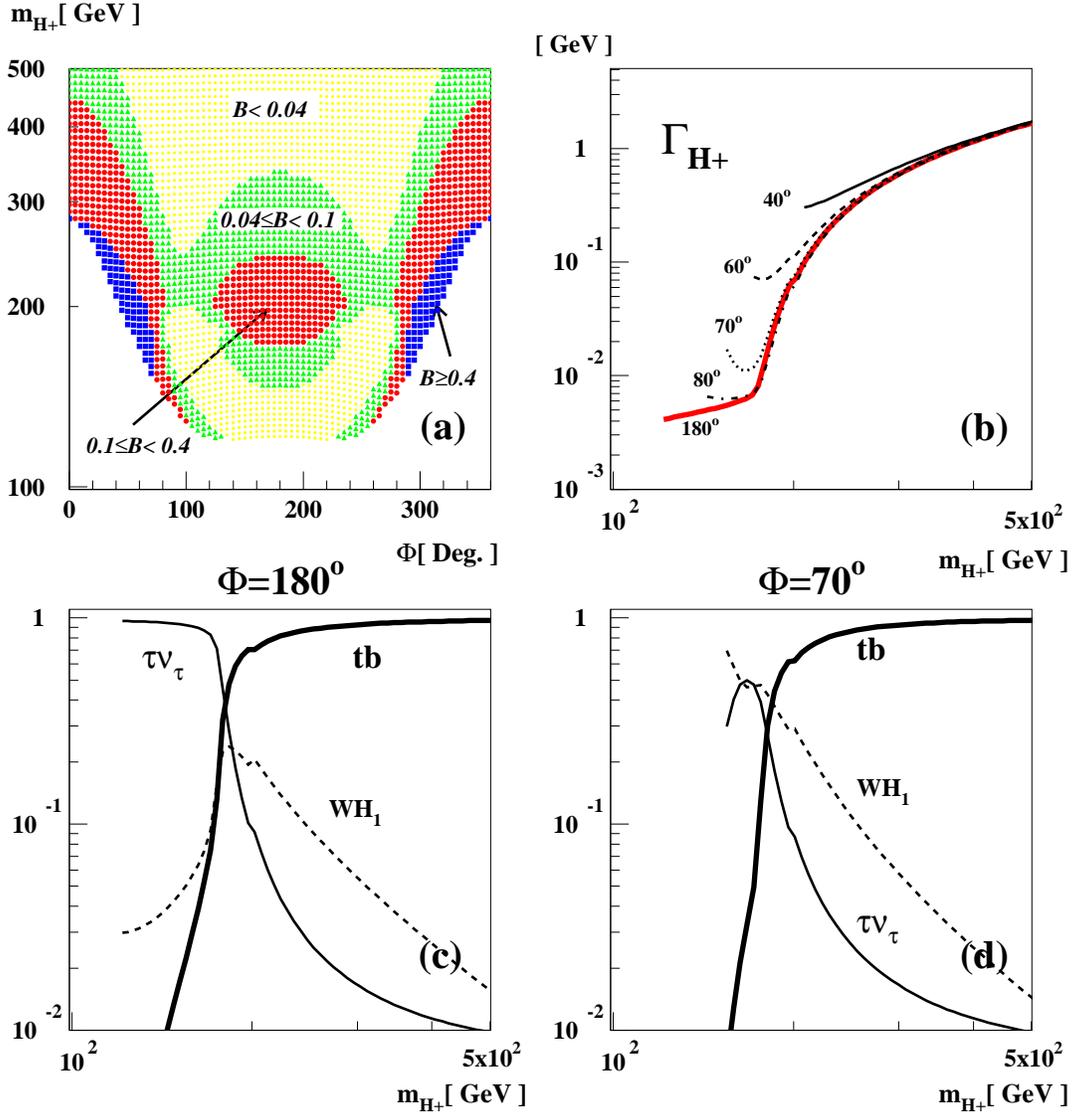,width=16cm,height=16cm}\hss}
\end{center}
\caption{The branching fractions of the charged Higgs boson and its total 
         decay width for $\tan\beta=4$ and the parameter set (\ref{eq:PARA});
         (a) the branching fraction ${\cal B}(H^+\rightarrow W^+H_1)$ on 
	 the plane of $m_{H^\pm}$ and $\Phi$, (b) the total decay width 
	 $\Gamma_{H^+}$ as a function of $m_{H^\pm}$ for five values of 
	 $\Phi$; $\Phi=180^{\rm o}$(thick line), 
	 $\Phi=80^{\rm o}$(dash--dotted line), $\Phi=70^{\rm o}$(dotted line), 
	 $\Phi=60^{\rm o}$(dashed line), and $\Phi=40^{\rm o}$(solid line), 
         (c) and (d) the branching fractions of the main charged--Higgs--boson 
	 decays as a function of $m_{H^\pm}$ for $\Phi=180^{\rm o}$ and 
	 $\Phi=70^{\rm o}$, respectively.}
\label{ch}
\end{figure}

For our numerical analysis of the charged--Higgs--boson decays, 
we take $\tan\beta=4$ and  the parameter set (\ref{eq:PARA}) as in the
analysis for the neutral Higgs--boson decays.
Fig.~\ref{ch}~(a) shows the branching fraction  
${\cal B}(H^+\rightarrow W^{+(*)} H_1)$ on the plane of $m_{H^\pm}$ and $\Phi$.
The allowed region is divided into 4 parts
according to the values of the branching fraction;
${\cal B}\geq 0.4$ (filled squares),
$0.1\leq {\cal B}<0.4$ (filled circles),
$0.04\leq {\cal B}<0.1$ (filled triangles), and
${\cal B}<0.04$ (dots).
The partial branching fraction ${\cal B}$ is enhanced with large CP
violating phase such that this mode could be significant for larger 
$m_{H^\pm}$ in the CP non--invariant case in contrast to the CP invariant 
case ($\Phi=180^{\rm o}$). 
The total decay width of the charged Higgs boson is shown in 
Fig.~\ref{ch}~(b) as a function of $m_{H^\pm}$. Each line corresponds to
$\Phi=180^{\rm o}$(thick line), $\Phi=80^{\rm o}$(dash--dotted line),
$\Phi=70^{\rm o}$(dotted line), $\Phi=60^{\rm o}$(dashed line), and
$\Phi=40^{\rm o}$(solid line). 
In the region with $\Phi \lesssim 80^{\rm o}$ and $m_{H^\pm}\lesssim 300$ GeV, 
the total decay width is also significantly affected by the CP phase $\Phi$. 
We show the
branching fractions of the charged Higgs boson in Fig.~\ref{ch}~(c) and (d)
as a function of $m_{H^\pm}$ for two representative values of $\Phi$:
$\Phi=180^{\rm o}$ and $\Phi=70^{\rm o}$. 
The thick solid line is for $H^+\rightarrow t^{(*)}\bar{b}$,
the solid line for $H^+\rightarrow \tau^+ \nu_\tau$, and
the dashed line for $H^+\rightarrow W^{+(*)}H_1$.
We find that the decay channel $H^+\rightarrow W^{+(*)} H_1$ is strongly
enhanced as the phase $\Phi$ differs from $180^{\rm o}$, and the channel
becomes the most dominant decay mode of the charged Higgs boson for
$m_{H^{\pm}}\lesssim 300$ GeV when the values of $\Phi$ are
less than $70^{\rm o}$. \\

To summarize, the charged Higgs boson decays into a $W$ boson and a neutral
Higgs boson are closely related to the neutral Higgs boson decays into gauge
boson pairs due to the orthogonality of the mixing
matrix $O$ in Eq. (\ref{OMIX}).  
As a result, the neutral and charged Higgs--boson decays, 
significantly affected by the phase $\Phi$, are complementary in confirming the
loop--induced CP violation in the MSSM Higgs sector.

%%%%%%%%%%%%%%%%%%%%%%%
\section{Conclusions}
\label{sec:conclude}
%%%%%%%%%%%%%%%%%%%%%%%
Based on the new calculation of the mass matrix of the neutral Higgs bosons
which is valid for any values of the relevant SUSY parameters, we have
re--evaluated the decays of the Higgs bosons in the MSSM with radiative
Higgs--sector CP violation. 
The present work extends the previous one \cite{CL1} 
by including all the neutral Higgs--boson decay modes containing
virtual gauge bosons and those into
top--squark pairs, implementing squark--loop contributions to the two--gluon
decay channel. We also study the decays of the charged Higgs boson in the
presence of the non--trivial CP--violating mixing.
In addition, we have discussed possible constraints from 
the electron EDM measurements on the CP phase $\Phi$. \\

We have found that the branching fractions of the main decay channels of the
lightest Higgs boson, whose mass is less than 120 GeV for $\tan\beta=4$, are
insensitive to the CP phase(s) and the Higgs boson mass.  
However, the sub--dominant decay channel $H_1\rightarrow V^{(*)}V^{(*)}$,
which is sensitive to the CP phase $\Phi$, can become
quite significant as the CP phase differs from $180^{\rm o}$. 
On the other hand, the decays of the heavy neutral Higgs bosons,
which are almost degenerate for $m_{H^{\pm}}>2\,m_Z$,
depend very strongly on the phase.
Below the thresholds of the decays into top--quark and top-squark pairs,
the main decay channels of the heavy Higgs bosons consist of
bottom--quark and tau--lepton pairs, (virtual) gauge bosons, 
Higgs and (virtual) $Z$ bosons, and Higgs bosons. The relative importance 
of those decay channels depends significantly on the size of the 
CP--violating mixing. The decay channels 
$H_i\rightarrow t\bar{t}\,, {\tilde t}_1\overline{{\tilde t}_1}
\,,{\tilde t}_1\overline{{\tilde t}_2}$ overwhelm all the other decay
channels once these channels are kinematically allowed.
The relative importance of these decay modes also depends crucially on the 
the CP phase(s) and the heavy Higgs--boson masses.  
Finally, the charged--Higgs--boson decays into a (virtual) $W$ boson and
a lightest neutral Higgs boson become dominant when the 
neutral Higgs--boson decays into gauge boson pairs are suppressed and the
charged--Higgs--boson mass is less than 300 GeV. \\

To conclude, the radiative Higgs sector CP violation induced from the
third--generation sfermion sectors could alter the patterns of the neutral
and charged Higgs--boson decays significantly from those in the CP invariant
theories.

%%%%%%%%%%%%%%%%%%%%%%%%%%%%%%%%
\section*{Acknowledgments}
%%%%%%%%%%%%%%%%%%%%%%%%%%%%%%%%

The authors thank A.~Pilaftsis for helpful discussion on the EDM constraints.
The work of S.Y.C.  was supported by a grant from the Korean Research 
Foundation Grant (KRF--2000--015--DS0009). The work of J.S.L. was supported 
by the Japan Society for the Promotion of Science (JSPS).

%\setcounter{equation}{0}
%\renewcommand{\theequation}{A\arabic{equation}}
%%%%%%%%%%%%%%%%%%%%%%%%%%%%
%\section*{Appendix}
%%%%%%%%%%%%%%%%%%%%%%%%%%%%

\vskip 1.5cm


\vfil\eject
%\end{multicols}

\end{document}